\documentclass[namedreferences]{solarphysics}

\usepackage[hyperref,optionalrh,showbiblabels]{spr-sola-addons} 
\usepackage{graphicx}        
\usepackage{color}           
\usepackage{breakurl}        

\renewcommand{\vec}[1]{{\mathbfit #1}}

\chardef\us=`\_

\newcommand{\omits}[1]{}

\begin{document}

\begin{article}
\begin{opening}

\title{Periods and damping rates of fast sausage oscillations in multi-shelled coronal loops}

\author[addressref={aff1}]{\inits{}\fnm{Shao-Xia}~\lnm{Chen}}
\author[addressref={aff1},corref,email={bbl@sdu.edu.cn}]{\inits{Bo}\fnm{Bo}~\lnm{Li}}
\author[addressref={aff1}]{\inits{}\fnm{Li-Dong}~\lnm{Xia}}
\author[addressref={aff1}]{\inits{}\fnm{Hui}~\lnm{Yu}}

\address[id=aff1]{Shandong Provincial Key Laboratory of Optical Astronomy and
Solar-Terrestrial Environment, and Institute of Space Sciences, Shandong University Weihai, Weihai 264209, China}

\runningauthor{Chen et al.}
\runningtitle{Sausage modes in multi-shelled coronal loops}

\begin{abstract}
Standing sausage modes are important in interpreting quasi-periodic pulsations in the lightcurves of solar flares.
Their periods and damping times play an important role in seismologically diagnosing
    key parameters like the magnetic field strength in regions where flare energy is released.
Usually such applications are based on theoretical results neglecting unresolved fine structures in magnetized loops.
However, the existence of fine structuring is suggested on both theoretical and observational grounds.
Adopting the framework of cold magnetohydrodynamics (MHD), we model coronal loops as magnetized cylinders
    with a transverse equilibrium density profile comprising a monolithic part and a modulation due to fine structuring
    in the form of concentric shells.
The equation governing the transverse velocity perturbation is solved with an initial-value-problem approach,
    and the effects of fine structuring on the periods $P$ and damping times $\tau$
    of global, leaky, standing sausage modes are examined.
A parameter study shows that fine structuring, be it periodically or randomly distributed,
    brings changes of only a few percent to $P$ and $\tau$ when there are more than about ten shells.
The monolithic part, its steepness in particular, plays a far more important role in
    determining $P$ and $\tau$.
We conclude that when measured values of $P$ and $\tau$ of sausage modes are used for seismological
    purposes, it is justified to use theoretical results where the effects due to fine structuring
    are neglected.
\end{abstract}
\end{opening}

\section{Introduction}
\label{sec_intro}
Low-frequency magnetohydrodynamic (MHD) waves and oscillations have been
    abundantly detected in the solar atmosphere with both spectroscopic
    and imaging instruments since the late 1990s
    with the advent of the TRACE, SOHO and \textit{Hinode} satellites.
This has led to the rapid development of
    coronal seismology,
    enabling the inference of atmospheric parameters
    difficult to measure directly~\citep[for recent reviews, see \textit{e.g.},][]{2005LRSP....2....3N,2012scsd.book.....S}.
To name but a few, seismological applications can offer such key information as
    the magnetic field strength in coronal loops~\citep[\textit{e.g.},][]{2001A&A...372L..53N}
    and above streamer stalks~\citep{2010ApJ...714..644C,2011ApJ...728..147C},
    the magnitude of field-aligned loop flows~\citep{2013ApJ...767..169L, 2014SoPh..289.1663C},
    as well as how the density is distributed transverse to coronal loops~\citep{2007A&A...463..333A,2008A&A...484..851G}.

In the context of coronal seismology, sausage waves
    were examined primarily in connection to
    their potential for interpreting second-scale quasi-periodic pulsations (QPPs)
    in flare lightcurves~\citep[see the review by][]{2009SSRv..149..119N}.
Two regimes of sausage waves are known to exist, depending on the wavenumber $k$ in the direction
    of magnetic structures hosting them \citep[\textit{e.g.},][]{1984ApJ...279..857R}.
When $k$ exceeds some critical $k_{\rm c}$, the trapped regime results,
    whereby the wave energy is well confined to magnetic structures.
If on the contrary, $k < k_{\rm c}$, the leaky regime results,
    and sausage waves experience apparent damping by emitting waves into
    the surrounding fluids.
For standing sausage modes, it is now well established that
    their period $P$ increases steadily with decreasing $k$
    until saturating {when $k$ is sufficiently small}.
Identically infinite in the trapped regime {for ideal MHD fluids}, the damping time $\tau$
    decreases with decreasing $k$ until a saturation value
    is reached {for sufficiently small $k$}~\citep[\textit{e.g.},][]{2007AstL...33..706K,2012ApJ...761..134N,2014ApJ...781...92V}.

Both the period $P$ and damping time $\tau$ of sausage modes are important
    from the seismological perspective due to their dependence on
    the atmospheric parameters.
Let $a$ {and $L$} denote the half-width {and length} of a coronal structure, {respectively.}
{In addition, let $v_{\rm Ai}$ ($v_{\rm Ae}$)}
    be the internal {(external)} Alfv\'en speed.
{For trapped modes, it is well known that the period $P$ is determined by $L/v_{\rm ph}$
    with the phase speed $v_{\rm ph}$ lying in the range between $v_{\rm Ai}$ and $v_{\rm Ae}$~\citep[\textit{e.g.},][]{2004ApJ...600..458A}.}
{On the other hand, for leaky modes} the period $P$ was found to depend primarily on the internal Alfv\'en
    transit time $a/v_{\rm Ai}$,
    while the ratio $\tau/P$ is primarily determined by the density contrast $\rho_{\rm i}/\rho_{\rm e}$
    between the structure and its surroundings~\citep[\textit{e.g.},][]{2007AstL...33..706K}.
For both $P$ and $\tau$, the detailed form, the steepness in particular, of the transverse density distribution
    can play a subtle role \citep{2012ApJ...761..134N, 2014A&A...567A..24H}.
{Regarding the applications of measurements of leaky sausage modes,}
    inferring $a/v_{\rm Ai}$ is important given the importance for inferring the coronal magnetic field strength.
The inference of $\rho_{\rm i}/\rho_{\rm e}$ and density profile steepness, on the other hand,
    is also important because these two parameters are key to determining
    the efficiency of such coronal heating mechanisms as
    phase-mixing \citep{1983A&A...117..220H} and resonant absorption \citep[see the review by][and references therein]{2011SSRv..158..289G}.

The current conclusions on the behavior of $P$ and $\tau$ of sausage modes
    are mainly based on theoretical and modeling efforts where the density structuring transverse to oscillating loops
    is either in a piecewise-constant (tophat) form \citep[\textit{e.g.},][]{1984ApJ...279..857R,2007AstL...33..706K}
    or described by some smooth functions \citep[\textit{e.g.},][]{2012ApJ...761..134N}.
However, both theoretical and observational studies indicate that coronal loops
    may not be regularly structured but contain unresolved, fine structuring.
For instance, density measurements~\citep{2009ApJ...694.1256T,2012ApJ...755L..33B},
    and temperature diagnostics~\citep[\textit{e.g.},][]{2008ApJ...686L.131W,2013ApJ...772L..19B}
    suggest that loops with apparent widths $\gtrsim 1000$~km
    are likely to comprise a multitude of fine structures, the scales of which may be
    down to $\lesssim 15$~km~\citep{2013A&A...556A.104P}.
On the theoretical side,
    numerical simulations on the temporal evolution of coronal
    loops~\citep[\textit{e.g.},][]{2000ApJ...528L..45R,2003ApJ...593.1174W, 2010ApJ...719..576G}
    indicate the need to invoke threads with widths $\sim 100$~km
    to better reproduce direct observables.
{Interestingly, further evidence showing the existence of fine structuring
    is found from a seismological standpoint by~\citet{2008A&A...491L...9V} when interpreting the CoMP measurements
    of propagating transverse waves.}
The increasing consensus on the existence of fine structuring in coronal structures
    has also stimulated the interest in examining the influence of fine structuring
    on coronal seismology~\citep[\textit{e.g.},][]{2001A&A...366..306M,2007SoPh..246..165P,2015ApJ...799..221Y}.
Regarding sausage modes, the study
    by~\citeauthor{2007SoPh..246..165P} (\citeyear{2007SoPh..246..165P}, hereafter PNA07) is of
    particular relevance since it focuses on how fine structuring influences the period $P$
    of trapped modes.
Modeling coronal loops as a magnetic slab with either periodical
    or random fine structuring,
    PNA07 find that $P$ of fundamental trapped modes is insensitive to
    fine structuring.

While the study by PNA07 is reassuring for seismological applications
    based on measured values of sausage periods,
    it is natural to ask:
    Does this conclusion hold also for magnetic cylinders?
    In addition, what will be the influence of fine structuring on
        the leaky modes, the damping time $\tau$ in particular?
These questions are addressed in the present study.
This manuscript is organized as follows.
In Section~\ref{sec_model} we present the necessary equations
    and our method of solution.
A parameter study is then presented in Section~\ref{sec_parameter}
    to examine in detail the changes to $P$ and $\tau$ due to fine structuring
    relative to the case where fine structuring is absent.
Section~\ref{sec_conclude} closes this manuscript with our summary
    and some concluding remarks.

\section{Model Description and Method of Solution}
\label{sec_model}

We work in the framework of cold (zero-beta) MHD, appropriate for the solar corona,
   and model coronal loops as straight cylinders
   whose axes coincide with the $z$-axis in a cylindrical coordinate system $(r, \theta, z)$.
The equilibrium magnetic field $\bar{\vec{B}}$ is uniform
   and also in the $z$-direction ($\bar{\vec{B}} = \bar{B}\hat{z}$).
The equilibrium density $\bar{\rho}$ is assumed to be a function
   of $r$ only.
In this regard, the modeled fine structures are concentric shells sharing the
    same axis as the cylinder.
Considering only fundamental standing sausage modes supported by
   loops of length $L$, we can express the radial component of the perturbed velocity
   as $\delta v_r (r, z, t) = v (r, t)\sin(kz)$ with the axial wavenumber $k$ being $\pi/L$.
It then follows from linearized, ideal, cold MHD equations that~\citep[see][]{2012ApJ...761..134N}
\begin{equation}
\label{eq_vr}
 \frac{\partial^2 v(r, t)}{\partial t^2}
    =v_{\rm A}^2(r)\left[\displaystyle\frac{\partial^2}{\partial r^2}
    +\displaystyle\frac{1}{r}
    \displaystyle\frac{\partial}{\partial r}-\left(k^2
    +\displaystyle\frac{1}{r^2}\right)\right]v (r, t)~,
\end{equation}
   where $v_{\rm A}(r)=\bar{B}/\sqrt{4\pi\bar{\rho}(r)}$ is the Alfv\'en speed.
The boundary conditions appropriate for sausage modes are
\begin{equation}\label{eq_boundary}
    \left.v(r,t)\right|_{r=0}=0,~~~~~~~
    \left.v(r,t)\right|_{r\to\infty}=0 .
\end{equation}
Furthermore, the following initial conditions (ICs)
\begin{equation}\label{eq_initial}
    \left.v (r,t)\right|_{t=0}=\displaystyle\frac{r}{1+r^2},~~~~~~~
    \left.\displaystyle\frac{\partial v (r,t)}{\partial t}\right|_{t=0}=0~
\end{equation}
    are adopted.
While in principle the ICs can be arbitrary, the form given in Equation~(\ref{eq_initial}) ensures that only waves with the simplest
    radial structure are primarily excited.

Similar to PNA07, the equilibrium density $\bar{\rho}(r)$
    consists of two parts, a background monolithic
    one ($\rho_{\rm mono}$) modulated by a distribution due to fine structuring ($\rho_{\rm FS}$).
In other words,
\begin{eqnarray}\label{eq_density}
  \bar{\rho}(r) &=& \rho_{\rm mono}(r)+\rho_{\rm FS}(r)~, \\
  \rho_{\rm mono}(r) &=& \rho_{\rm e}+\left(\rho_{\rm i}-\rho_{\rm e}\right)f(r)~, \\
  \rho_{\rm FS}(r) &=& \left(\rho_{\rm i}-\rho_{\rm
  e}\right)g(r)f(r)~ .
\end{eqnarray}
The function $f(r)$ decreases smoothly from unity at the cylinder axis ($r=0$)
   to zero at infinity, ensuring that
   $\rho_{\rm mono}(r)$ decreases smoothly from $\rho_{\rm i}$ at $r=0$
   to $\rho_{\rm e}$ at infinity.
On the other hand, $g(r)$, which describes the fine structures in the form of concentric shells,
   cannot exceed unity to avoid negative values of $\bar{\rho}$.
To maximize the effects due to fine structuring,
   we set the maximum of $g(r)$ to be unity.

With the boundary (Equation~(\ref{eq_boundary})) and initial (Equation~(\ref{eq_initial})) conditions, 
   Equation~(\ref{eq_vr}) can be evolved once $f(r)$ and $g(r)$ are specified.
In practice, we solve Equation~(\ref{eq_vr}) with a finite-difference (FD) scheme second-order accurate
   in both time and space.
To save computational time, a non-uniform computational grid $\{r_i, i=1, 2, \cdots, I\}$ is adopted
   with $r_1 = 0$ and $r_I = 1000a$, where $a$ is the cylinder radius.
The grid spacing $\{\Delta r_i = r_{i+1}-r_i, i= 1, 2, \cdots, I-1\}$ is identically $0.001a$ for $r\le 4a$.
Then $\Delta r_i$ increases in the manner $\Delta r_{i+1} = 1.025\Delta r_i$ until
   $\Delta r_i$ reaches $\sqrt{\rho_{\rm i}/\rho_{\rm e}}\times \Delta r_1$.
From there on $\Delta r_i$ remains uniform again.
To ensure numerical stability, a uniform timestep $\Delta t = 0.8 \Delta r_{\rm min}/v_{\rm A, max}$
   is set according to the Courant condition, where $\Delta r_{\rm min}$
   ($v_{\rm A, max}$) denotes
   the minimal (maximal) value of $\{\Delta r_i\}$ ($\{v_{{\rm A},i} = \bar{B}/\sqrt{4\pi\bar{\rho}(r_i)}\}$).
In response to the initial condition (Equation~(\ref{eq_initial})), disturbances are generated and propagate away from the cylinder.
However, as shown in \citet{2012ApJ...761..134N}, two regimes can be readily distinguished
   if one follows the temporal evolution of the perturbation at, say, $r=a$.
For $k=\pi/L$ larger (smaller) than some critical value,
   this $v(a, t)$ evolves into a harmonic (decaying harmonic) form,
   corresponding to the well-known trapped (leaky) regime.
Numerically fitting $v(a, t)$ with a sinusoidal (exponentially decaying sinusoidal) function
   then yields the period $P$ ($P$ together with the damping time $\tau$)
   for trapped (leaky) modes.
It suffices to note here that $P$ and $\tau$ depend only on the combination
   $[f(r), g(r); L/a, \rho_{\rm i}/\rho_{\rm e}]$
   when $P$ and $\tau$ are measured in units of the internal Alfv\'en transit time
   $a/v_{\rm Ai}$ with $v_{\rm Ai}$ being $\bar{B}/\sqrt{4\pi\rho_{\rm i}}$.

A number of measures are taken to ensure the accuracy of the numerical results.
First, a grid convergence test is made for
   a considerable fraction of the computations, whereby the numerical results
   do not show any appreciable change when the grid spacing is halved.
Second, we work only with the part of the $v(a, t)$ signal when traveling disturbances
   have not reached $r_I$, meaning that the signal is not contaminated
   from disturbances reflected off this outer boundary.
Third, we have tested other forms of the initial condition to make sure that
   the derived values of $P$ and $\tau$ do not depend on this choice.
However, since we are interested only in the lowest order modes,
   we choose not to use a too localized $v (r, t=0)$ since this excites higher-order sausage modes
   as well.
Forth, for a number of $[f(r), g(r)]$, we also compute $P$ for trapped modes
   by formulating Equation~(\ref{eq_vr}) into an eigen-value problem and then numerically solving
   it with a MATLAB boundary-value-problem solver BVPSUITE in its eigen-value mode
(See \citeauthor{2009AIPC.1168...39K}~\citeyear{2009AIPC.1168...39K} for a description of the solver,
   and \citeauthor{2014A&A...568A..31L}~\citeyear{2014A&A...568A..31L} for its recent application to sausage modes).
The values of $P$ thus found agree remarkably well
   with those found with our initial-value-problem
   approach (see Figure~\ref{fig_vsk}).

\section{Parameter study}
\label{sec_parameter}

In this section, we perform a parameter study to see how the fine
   structuring of a multi-shelled loop affects
   the period $P$ and damping time $\tau$ of the
   global sausage modes.
Given that $[f(r), g(r); L/a, \rho_{\rm i}/\rho_{\rm e}]$ constitutes too large a parameter space
   to exhaust, we choose to fix the density contrast $\rho_{\rm i}/\rho_{\rm e}$ to be 50,
   in accordance with the value suggested by the event reported in~\citet{2003A&A...412L...7N}.
In addition, we fix the form of $f(r)$ to be
\begin{eqnarray}
  \label{eq_fr}
  f(r) = \exp\left[-\left(\frac{r}{a}\right)^p\right],  \hspace{0.2cm} p>1.
\end{eqnarray}
Obviously, $\rho_{\rm mono}(r)$ is increasingly steep when $p$ increases.
This choice of $f(r)$ is similar to the symmetric Epstein profile adopted in PNA07 when $p\approx 2$,
   however, a variable $p$ also allows us to see the effects of the steepness of
   the monolithic density profile.
Unless otherwise specified, we fix $p$ at $2$.

\subsection{Periodic Fine Structuring}
\label{sec_sub_periodicFS}

\begin{figure}
\centerline{\includegraphics[width=0.9\textwidth]{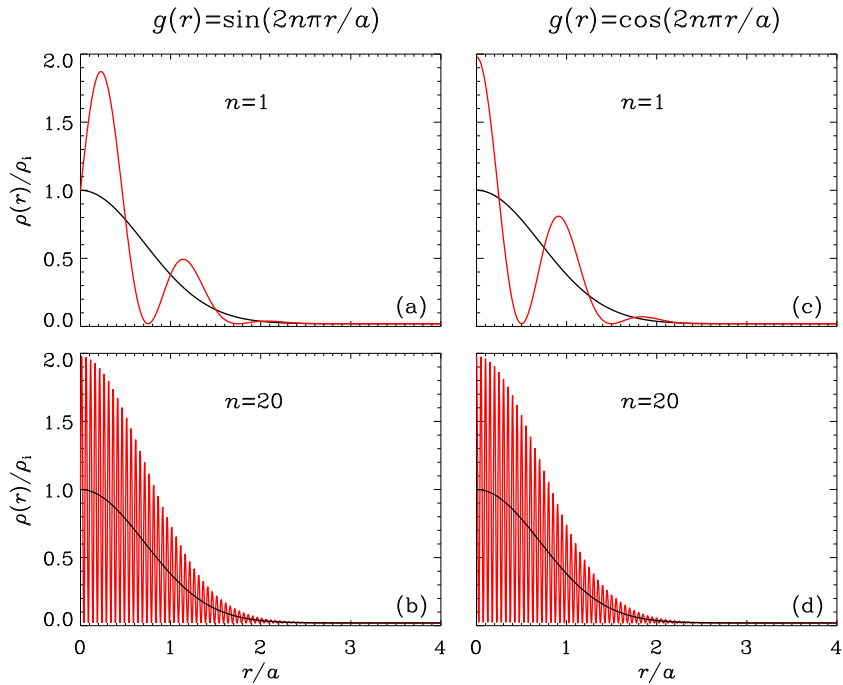}}
 \caption{Equilibrium density profiles ($\bar{\rho}(r)$) with periodic fine structuring.
 The red curves in the left (right) column are for $\bar{\rho}(r)$ in which
     fine structuring is given by sinusoidal (cosinusoidal) forms.
 The black curves present the monolithic component, to which a modulation due to fine structuring is added.
 Furthermore, two values for the number of fine structures, $1$ (the upper row) and $20$ (lower),
     are adopted for illustration purposes.
 }
 \label{fig_sin}
\end{figure}

Let us start with examining fine structuring of the form,
\begin{equation}
  g(r)=\left\{\begin{array}{c}
  \sin\displaystyle\frac{2n\pi r}{a}~, \\
  [0.3cm] \cos\displaystyle\frac{2n\pi r}{a}~,
\end{array}
\right.
\end{equation}
   with the integer $n$ representing the number of shells in a loop.
Figure~\ref{fig_sin} shows the equilibrium density profile $\bar{\rho}$
   as a function of $r$, with sinusoidal (cosinusoidal) modulations given by the red curves
   in the left (right) column.
Two values of $n$, $1$ (the upper row) and $20$ (lower),
   are adopted for illustration purposes.
For comparison, the $\bar{\rho}$ profile without fine structuring is given by
   the black solid lines.

\begin{figure}
\centerline{\includegraphics[width=0.9\textwidth]{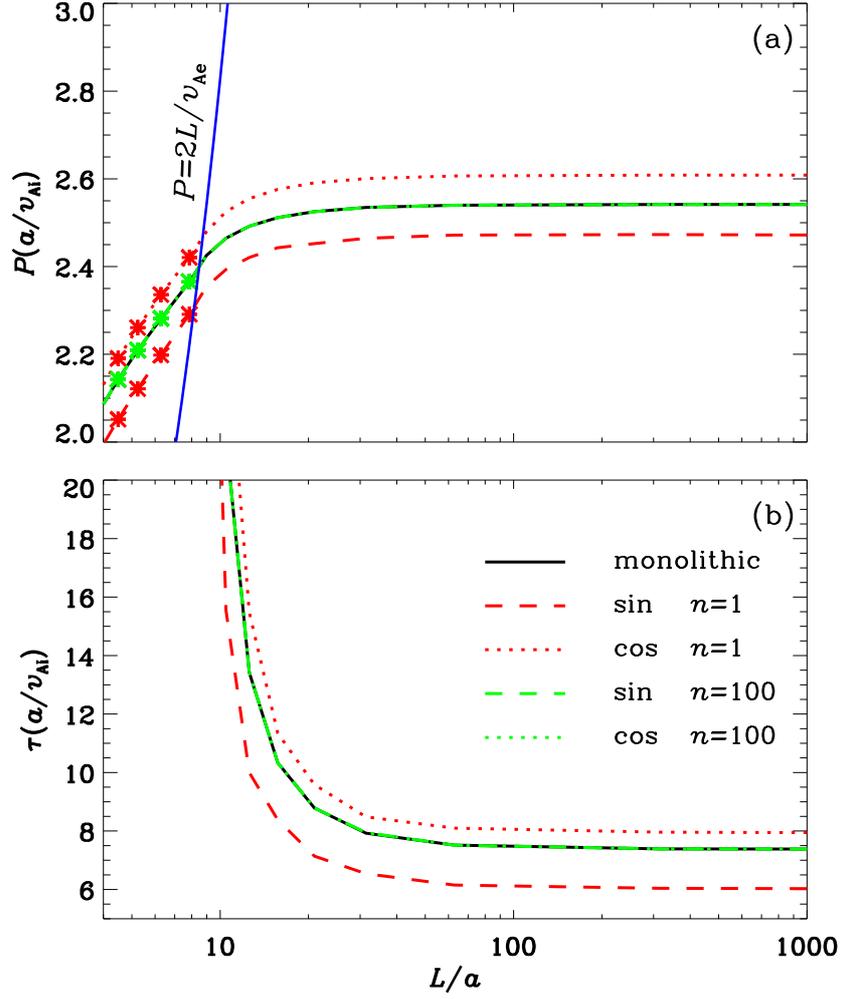}}
 \caption{(a) Period $P$ and (b) damping time $\tau$ of sausage modes supported by loops with periodic fine structuring.
 Here $P$ and $\tau$ are displayed as functions of loop length $L$.
 The black curves represent the cases where fine structuring is absent.
 The dashed (dotted) curves are for the results where fine structuring is in a sinusoidal (cosinusoidal) form.
 Two values for the number of fine structures, $1$ (the red curves) and $100$ (green),
     are examined.
 In (a), the blue solid line separates trapped (to its left) from leaky (right) waves.
 Furthermore, the asterisks present the periods of trapped modes derived by
    solving the problem from an eigen-value-problem perspective (see text for details).
}
\label{fig_vsk}
\end{figure}

Figure~\ref{fig_vsk} shows the dependence on the length-to-radius ratio $L/a$
   of the period $P$ and damping time $\tau$.
As labeled in Figure~\ref{fig_vsk}b, the dashed (dotted) curves correspond to the choice
   of $g(r)$ as a sinusoidal (cosinusoidal) function,
   and the red (green) curves represent the cases where $n=1$ ($n=100$).
We note that with the asymptotic Alfv\'en speed $v_{\rm Ae}$ defined
   as $\bar{B}/\sqrt{4\pi\rho_{\rm e}}$, the maximal period that trapped modes attain
   is $2 L/v_{\rm Ae}$.
This is given in Figure~\ref{fig_vsk}a by the blue solid line,
   which separates the trapped (to its left, where $\tau$ is identically infinite)
   from leaky (to its right) regimes.
For comparison, the black curves represent
   the cases where fine structuring is absent (hardly discernible from the green curves though).
The curves are found from our fitting procedure, whereas in Figure~\ref{fig_vsk}a the asterisks on the left of
   the blue line are found by formulating Equation~~(\ref{eq_vr}) into an eigen-value problem
   and then solving it with BVPSUITE.
Obviously, the periods $P$ found with BVPSUITE agree remarkably well with those
   from the fitting procedure.
Figure~\ref{fig_vsk} indicates that while differing in details,
   the overall tendency for $P$ ($\tau$) to increase (decrease)
   with $L/a$ is seen for all the equilibrium density profiles considered.
In particular, regardless of the profiles, at large values of $L/a$,
   both $P$ and $\tau$ saturate at some asymptotic values.
In addition, while for $n=1$ (the red curves) the $P$ and $\tau$ values
   show some appreciable difference from the case without fine structuring (the black curves),
   for $n=100$ (the green curves) the values of $P$ and $\tau$ can hardly be told apart from
   the black curves.

\begin{figure}
\centerline{\includegraphics[width=0.9\textwidth]{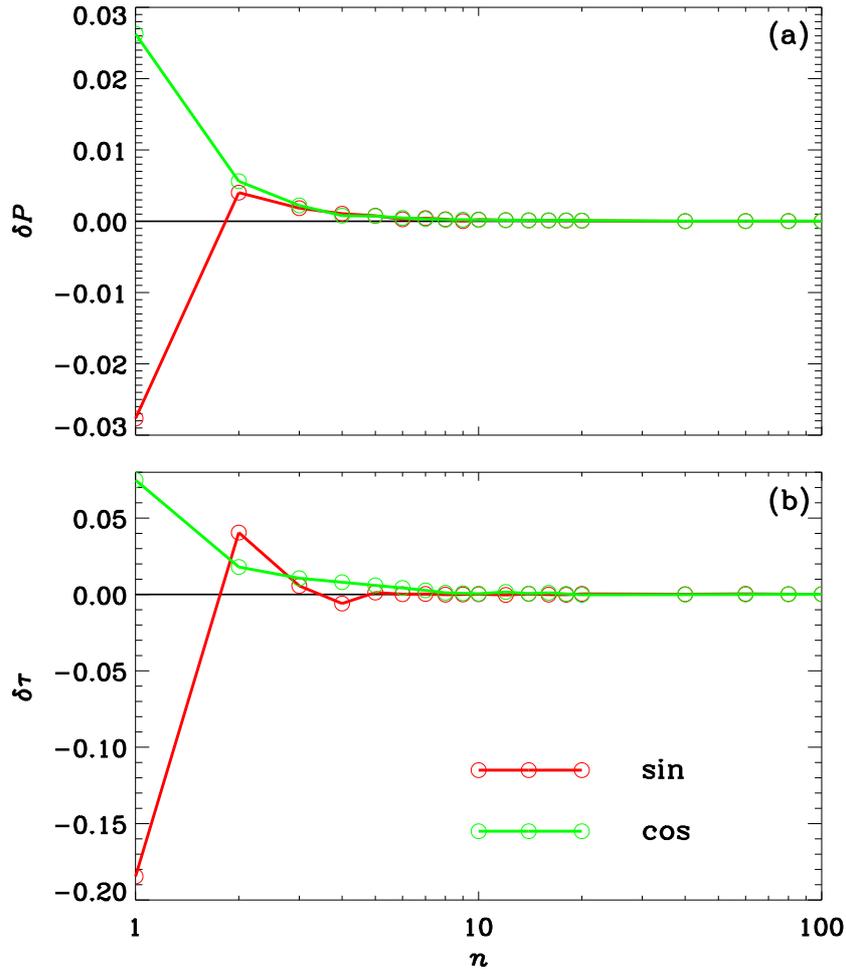}}
 \caption{Dependence on the number of shells ($n$) of $\delta P$ and $\delta \tau$,
     the fractional changes to the periods and damping times of sausage
     modes supported by thin loops with periodic fine structuring.
 The red (green) curves represent the cases where fine structuring is
     in a sinusoidal (cosinusoidal) form.
}
\label{fig_vsn}
\end{figure}

From the computations with the two extreme values of $n$ ($n=1$ and $100$),
   one may reasonably conjecture that the larger $n$ is, the less the effect
   associated with the periodic fine structuring.
To see whether this is true, we employ the fact that neither the period $P$
   nor damping time $\tau$ depends on $L/a$ when $L/a$ is sufficiently large.
Let $P_{\rm s}$ and $\tau_{\rm s}$ denote the asymptotic values that $P$ and $\tau$ attain,
   respectively.
The $n$-dependence can then be brought out by examining how varying $n$ changes
   $P_{\rm s}$ and $\tau_{\rm s}$ relative to the cases where
   fine structuring is absent (denoted by the subscript mono).
These are defined as
\begin{eqnarray}
\label{relative}
\delta P   \equiv \frac{P_{\rm s}-P_{\rm mono,s}}{P_{\rm mono,s}}, \hspace{0.2cm}
\delta \tau\equiv \frac{\tau_{\rm s}-\tau_{\rm mono,s}}{\tau_{\rm mono,s}}~.
\end{eqnarray}
Figure~\ref{fig_vsn} presents (a) $\delta P$
   and (b) $\delta \tau$ as a function of $n$
   for $g(r)$ in the sinusoidal (cosinusoidal) form as given by the red (green) curves.
One can see that indeed the fractional variations in $P$ and $\tau$
   tend to decrease in magnitude with increasing $n$.
However, even in the cases where $n=1$, $|\delta P|$ is no larger than $3\%$,
   and $|\delta \tau|$ remains less than $20\%$, with the most prominent changes found
   for $|\delta \tau|$ when a sinusoidal form is adopted for $g(r)$
   (the red curve in Figure~\ref{fig_vsn}b).
This happens despite that the equilibrium density profiles with fine structuring are remarkably different
   from the one without it (see Figures~\ref{fig_sin}a and \ref{fig_sin}c).
The reason for the tendency for the periodic fine structuring to
have less effect
   with increasing number of shells is related to the transverse profile of
   the corresponding perturbations.
While not shown here, these profiles have spatial scales of the order of the cylinder radius.
With increasing $n$, the spatial scale ($\sim a/n$) of the transverse density profile
   decreases (\textit{e.g.}, compare Figures~\ref{fig_sin}a with \ref{fig_sin}b).
As one expects that the effect of fine structuring maximizes when
the two scales are comparable,
   for large values of $n$, the density fine structuring will have too small a transverse
   spatial scale to influence $P$ or $\tau$.

\subsection{Random Fine Structuring}
Compared with the periodic form, a step closer to reality will be
   a random distribution of the fine structures (see also PNA07).
This leads us to examine
\begin{equation}\label{eq_random}
g(r)=\frac{R}{R_{\rm max}}, \hspace{0.2cm}
   R = \sum^N_{n=1}A_n\sin\left(\displaystyle\frac{2n\pi
    r}{a}+\phi_n\right)~,
\end{equation}
where $A_n$ and $\phi_n$ are two independent arrays of random numbers
   in the ranges $[0,~1]$ and $[0,~2\pi]$, respectively.
The denominator $R_{\rm max}$ represents the maximum that $R$ attains,
   thereby ensuring that $g(r) \le 1$.
Moreover, $N$ stands for the number of harmonics that take part in
   the fine structuring, and can roughly represent the number of
   randomly distributed concentric shells.
For an illustration of the equilibrium density profile
   associated with such a $g(r)$, see the green curve in Figure~\ref{fig_profilep}b where $N$ is set to be $60$.

\begin{figure}
\centerline{\includegraphics[width=0.9\textwidth]{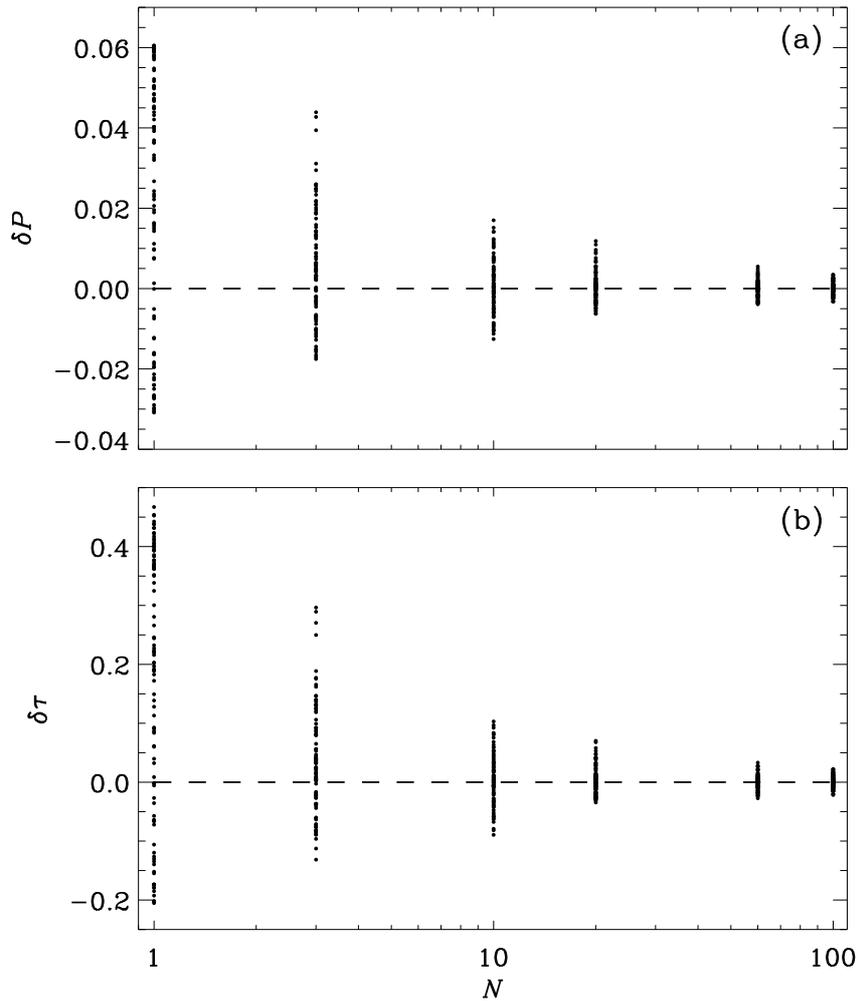}}
 \caption{Dependence on $N$ of $\delta P$ and $\delta \tau$, the fractional changes to the periods
     and damping times of sausage modes supported by thin loops with random fine structuring.
 Here $N$ represents the number of harmonics that enter into Equation~(\ref{eq_random}),
     and can be seen as the number of randomly distributed shells.
 At every value of $N$, each dot represents one of the $100$ computed realizations of the random structuring.
}
 \label{fig_vsN}
\end{figure}

Obviously, with increasing $N$, the relative importance of the harmonics with lower $n$
   decreases.
From Figure~\ref{fig_vsn} we know that for single harmonics, the
changes to both the period $P$ and damping time $\tau$
   decrease with increasing $n$.
One then expects that as $N$ increases, the influence of random
   fine structuring decreases.
It turns out that this is indeed the case if one examines Figure~\ref{fig_vsN},
    where the influence of fine structuring is represented by $\delta P$ and $\delta \tau$,
   the fractional changes to the saturation values $P_{\rm s}$ and $\tau_{\rm s}$
   relative to the case where
   fine structuring is absent.
Each dot for a given $N$ represents a realization of the random function $R$,
   and in total 100 realizations are computed.
Figure~\ref{fig_vsN}a indicates that $|\delta P|$ is
consistently smaller than $6\%$, meaning that the period of
   the fundamental sausage mode is insensitive to fine structuring.
Rather, it is primarily determined by the monolithic part of the equilibrium density profile.
Note that this conclusion was established in PNA07 for trapped modes supported by magnetic slabs, and here
    we have shown that it holds for leaky modes in magnetic cylinders as well.
On the other hand, Figure~\ref{fig_vsN}b shows that the effect due to
   fine structuring may be important
   in determining the damping time: for some realizations $\delta \tau$ can reach $47\%$.
However, $\delta \tau$ becomes less than $10\%$ when $N \gtrsim 10$.
Let us consider what this implies for coronal seismology, assuming that only $P$ and $\tau$
are available as observables
   and as generally accepted, loops consist of a substantial number of fine structures.
The negative side is that the information on fine structuring can not be probed since
the changes to $P$
   and $\tau$ due to fine structuring may be well below measurement uncertainties.
However, the positive side is that, when inverting the observables $P$ and $\tau$ to infer, say,
   the internal Alfv\'en transit time, it suffices to use theoretical
   results accounting for only the simpler, monolithic equilibrium density profile.

\subsection{Effect of the Monolithic Density Profile}

\begin{figure}
\centerline{\includegraphics[width=0.9\textwidth]{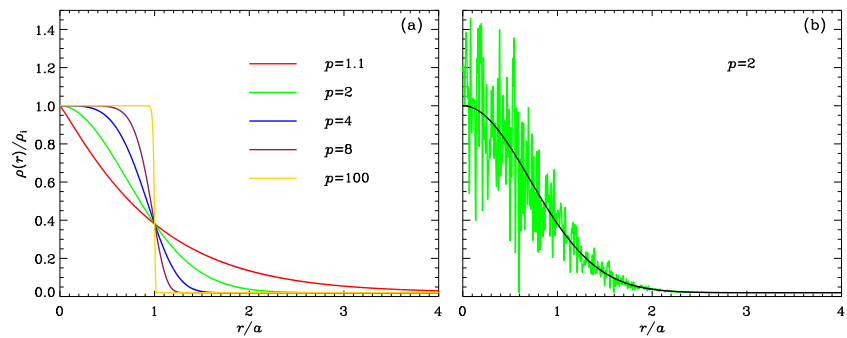}}
 \caption{Equilibrium density profiles (a) without and (b) with random fine structuring.
 In (a), a series of values for $p$, the steepness parameter, is shown by curves
    with different colors.
 In (b), the black curve represents the monolithic component of the density profile
    given by the green curve, which includes contributions from $60$ harmonics (see Equation~(\ref{eq_random})).
 For this monolithic component, a value of $p=2$ is adopted.
 }
 \label{fig_profilep}
\end{figure}

So far we have fixed the steepness index $p$ to be $2$ when specifying $f(r)$.
One may naturally question what would happen if $p$ is varied?
To start, let us illustrate how $p$ influences the monolithic part of the transverse density profile
   as given in Figure~\ref{fig_profilep}a.
As expected, the monolithic density profile becomes steeper as $p$ increases.

\begin{figure}
\centerline{\includegraphics[width=0.8\textwidth]{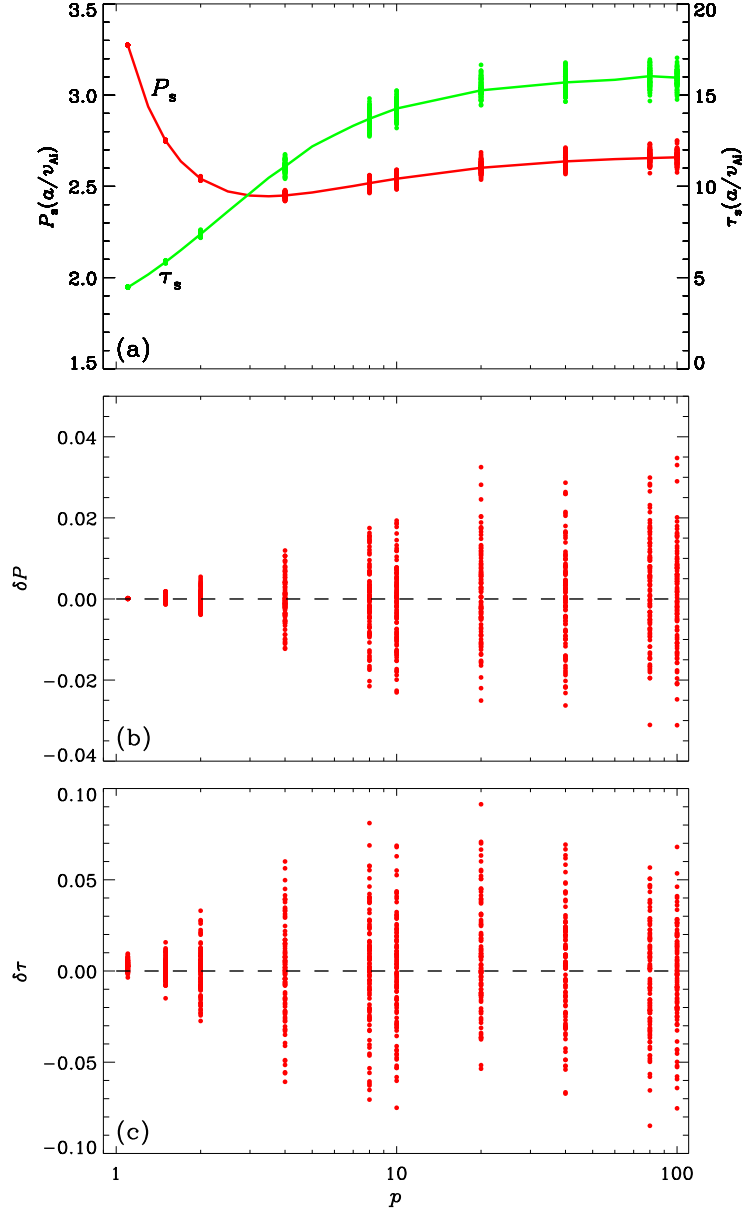}}
 \caption{
 Dependence on $p$, the density profile steepness,
    of the periods $P_{\rm s}$ and damping times $\tau_{\rm s}$ of
    sausage modes supported by thin loops with random fine structuring.
 In panel (a), the absolute values of $P_{\rm s}$ and $\tau_{\rm s}$ are given,
    while their fractional changes relative to the cases where fine structuring is absent
    are given in panels (b) and (c), respectively.
 The random fine structuring includes $60$ harmonics.
 For every $p$ value, each dot represents one of the $100$ realizations of the random fine structuring.
}
\label{fig_vsp}
\end{figure}

Figure~\ref{fig_vsp}a shows the variation with $p$ of $P_s$ (the red dots) and $\tau_s$ (green),
   the saturation values of the period and damping time.
A random structuring as given by Equation~(\ref{eq_random}) with $N=60$ is adopted.
Any dot at each value of $p$ represents one of the 100 realizations of the random function $R$,
   while the curves correspond to the case where random structuring is absent.
The relative changes $\delta P$ and $\delta \tau$ are then presented in Figures~\ref{fig_vsp}b and \ref{fig_vsp}c, respectively.
One can see that the effect of fine structuring tends to increase
   with increasing $p$.
In fact, for $p=1.1$, the dots in Figure~\ref{fig_vsp}a hardly show
any appreciable deviation from the curves,
   a result naturally expected since Figures~\ref{fig_vsp}b and \ref{fig_vsp}c
   indicate that $|\delta P|$ and $|\delta \tau|$
   are no larger than $1\%$.
In contrast, for $p=100$, Figures~\ref{fig_vsp}b and \ref{fig_vsp}c
    indicate that $|\delta P|$ and $|\delta \tau|$ may be up to
    $3.48\%$ and $7.53\%$.
However, these values at a given $p$ are marginal to say the most when compared with the changes to $P_s$ and $\tau_s$ due to variations in
   the steepness index $p$.
For instance, Figure~\ref{fig_vsp}a shows that in the case without fine structuring,
   $\tau_s$ increases monotonically with $p$, attaining $4.46~a/v_{\rm Ai}$ ($16~a/v_{\rm Ai}$) when $p=1.1$ ($p=100$).
In relative terms, $\tau_s$ with $p=100$ is $3.58$ times that in the case where $p=1.1$.
From this comparison we conclude that at least for the density profiles we explored,
   compared with the fine structures,
   the monolithic part plays a far more important role in determining the period
   and damping time.

\section{Conclusions}
\label{sec_conclude}

On both theoretical and observational grounds, magnetic loops in the solar corona are suggested to comprise
    a multitude of fine structures with transverse scales much shorter than loop widths.
From the coronal seismology perspective, while the effects of fine structuring in the form of multilayered slabs
    on the periods of trapped, standing sausage
    modes were shown to be at most marginal~\citep{2007SoPh..246..165P},
    it remains to be seen whether the conclusion also holds for coronal cylinders and
    for leaky modes.
Working in the framework of cold magnetohydrodynamics (MHD), we model coronal loops as magnetized cylinders
    with a transverse equilibrium density profile
    comprising a monolithic part and a modulation due to fine structuring.
This kind of fine structuring can be thought of as constituting concentric shells sharing
    the same axis with the cylinder itself.
We focused on fundamental modes with the simplest transverse structure, namely,
    the lowest order sausage modes without extra nodes between
    the two ends of the cylinder.
The equation governing the transverse velocity perturbation was solved with an initial-value-problem approach,
    thereby the period $P$ and damping time $\tau$ were derived by numerically fitting
    the signals at some given location away from the cylinder axis.
Both $P$ and $\tau$ were shown to saturate at some asymptotic values in the slender-cylinder limit.
This enabled us to adopt $\delta P$ and $\delta \tau$, the fractional changes to these saturation values relative to the cases where fine structuring is absent,
    as indicators of the effects associated with the presence of fine structuring.

Starting with an examination of periodical fine structures,
    we showed that as the number of shells increases, the effects on the period and damping time of leaky sausage modes
    decrease.
This is attributed to the disparity between the characteristic transverse scale of the velocity perturbation
    and the increasingly small scale of the fine structures with increasing shell number.
When the shell number exceeds $\sim 2$, both $\delta P$ and $\delta \tau$ amount to a few percent.
Going a step closer to reality, we examined whether random fine structuring can have a more prominent effect
    in determining $P$ and $\tau$.
While $\delta P$ and $\delta \tau$ are indeed larger for some realizations of the fine structuring
    than in the cases with periodic structuring,
    when the number of fine structures (roughly represented by the number of harmonics that enter into
    the random structuring) exceeds $\sim 10$, neither $\delta P$ nor $\delta \tau$
    is in excess of $10\%$.
We showed by varying the steepness of the monolithic part of the density profile
    that this change may bring forth changes to $P$ and $\tau$ by a factor of several,
    far more prominent than the effects due to fine structuring.

Our findings can be considered positive news for seismological applications of
    the period and damping time of leaky sausage modes using
      theoretically~\citep[\textit{e.g.},][]{2007AstL...33..706K, 2014ApJ...781...92V}
      or numerically~\citep[\textit{e.g.},][]{2012ApJ...761..134N}
    derived $P$ and $\tau$ where fine structuring is neglected.
However, formulating fine structures as concentric shells should be considered only
    a preliminary step closer to reality.
Fine structures in realistic coronal loops may be organized as randomly distributed strands,
    thereby requiring the problem be formulated as a two-dimensional one
    in the plane transverse to coronal cylinders~\citep[\textit{e.g.},][]{2010ApJ...716.1371L}.
Furthermore, to simplify our treatment, we have neglected the finite plasma beta
    and the longitudinal variation of the equilibrium magnetic field or density.
These two factors are unlikely to be important,
    though~\citep[see][]{2009A&A...494.1119P, 2009A&A...503..569I}.

\begin{acks}
This research is supported by the 973 program 2012CB825601, National Natural Science Foundation of China
    (41174154, 41274176, 41274178, and 41474149),
    and by the Provincial Natural Science Foundation of Shandong via Grant JQ201212.
\end{acks}





\bibliographystyle{spr-mp-sola}
\bibliography{chen}

\IfFileExists{\jobname.bbl}{} {\typeout{}
\typeout{****************************************************}
\typeout{****************************************************}
\typeout{** Please run "bibtex \jobname" to obtain} \typeout{**
the bibliography and then re-run LaTeX} \typeout{** twice to fix
the references !}
\typeout{****************************************************}
\typeout{****************************************************}
\typeout{}}

\end{article}

\end{document}